\documentclass[12pt, english]{article}

\usepackage{amsmath}
\usepackage[hidelinks]{hyperref}
\usepackage{cleveref}
\usepackage{multicol}
\usepackage{amssymb}
\usepackage{csquotes}
\usepackage{babel}
\usepackage{booktabs}
\usepackage[shortlabels]{enumitem}
\usepackage{hhline}
\usepackage{multirow}
\usepackage{makecell}
\usepackage{diagbox}
\usepackage[toc, page]{appendix}
\usepackage{rotating}
\usepackage{color}
\usepackage{float,lscape}
\usepackage{graphicx}
\graphicspath{{./figures/}}

\usepackage[backend=biber,style=numeric,sorting=nyt]{biblatex}
\newcommand{\mathval}[1]{$#1$}
\newcommand{\best}[1]{\color{green}{$#1$}}
\newcommand{\stdcell}[2]{\makecell{$#1$\\ $\pm#2$}}
\newcommand{\beststdcell}[2]{\color{green}{\makecell{$#1$\\ $\pm#2$}}}
\newcommand{\R}{{\mathbb{R}}}
\newcommand{\ignore}[1]{}

\begin{document}

\title{Revisiting information cascades in online social networks}
\author{
    Michael Sidorov\\
    Department of Communication Systems Engineering\\Ben Gurion University of the Negev\\Be'er Sheba, Israel\\
    sidorov@post.bgu.ac.il\\
    \and
    Dan Vilenchik\\
    Department of Communication Systems Engineering\\Ben Gurion University of the Negev\\Be'er Sheba, Israel\\
    vilenchi@bgu.ac.il
}
\date{Mar, 2021}
\maketitle

\begin{abstract}
It's by now folklore that to understand the activity pattern of a user in an online social network (OSN) platform, one needs to look at his friends or the ones he follows. The common perception is that these friends exert influence on the user, effecting his decision whether to re-share content or not. Hinging upon this intuition, a variety of models were developed to predict how information propagates in OSN, similar to the way infection spreads in the population. In this paper, we revisit this world view and arrive at new conclusions. Given a set of users $V$, we study the task of predicting whether a user $u \in V$ will re-share content by some $v \in V$ at the following time window given the activity of all the users in $V$ in the previous time window. We design several algorithms for this task, ranging from a simple greedy algorithm that only learns $u$'s conditional probability distribution, ignoring the rest of $V$, to a convolutional neural network-based algorithm that receives the activity of all of $V$, but does not receive explicitly the social link structure. We tested our algorithms on four datasets that we collected from Twitter, each revolving around a different popular topic in 2020. The best performance, average F1-score of 0.86 over the four datasets, was achieved by the convolutional neural network. The simple, social-link ignorant, algorithm achieved an average F1-score of 0.78. 

\end{abstract}

\section{Introduction}
The propagation of ideas and innovations has
led to political, cultural and economic  changes. Social
scientists found that ideas tend to flow along social links, in a similar manner that an epidemic infects a population~ \cite{delre2010will,difOfInov}. Understanding how information propagates in online social network (OSN) platforms has significant real-world impact, from the political stability of states to marketing. Significant efforts have been invested in predicting various properties of information cascades, such as size \cite{lerman2010information}, temporal growth \cite{li2017deepcas}, and virality \cite{cheng14}.

Researchers mainly studied transient-copy propagation protocols (a term coined in \cite{ChengKLLSSA18}), where content is simply replicated in a network, like a re-tweet (contrasted with nomination cascades for example, such as the famous ALS ice water bucket challenge). 

It is tempting to think that information propagates in OSNs according to epidemiological models such as Bass \cite{bass1969new} or SIR \cite{kermack1927contribution}. However several ``cautionary tales" were told by various researchers regarding the validity of this image. For example, one important assumption in epidemiological models is that all newly infected individuals arise from the susceptible set (neighbors of infectious people). However, mass media marketing efforts rely on a "broadcast" mechanism, where a large number of individuals can receive the information directly from the same source, which need not be their neighbor. Indeed, \cite{cha2009measurement} who studied information cascades in Flickr, found that out of 10 million total favorite markings, 47\%  were propagated not through the social links, but via other broadcasting mechanisms that Flickr employs. Also \cite{galuba2010outtweeting} found that  33\% of the retweets in their dataset credit users that the retweeters do not follow. 

Furthermore, unlike the epidemiological picture of a disease spreading in waves, it was observed  that most cascades are extremely shallow and wide. For example, \cite{goel2016structural} found that the average size of the diffusion trees in Twitter is 1.3, and that the vast majority contain only 1 node. This phenomenon is also observed in other platforms such as Digg and Flickr \cite{cha2009measurement,lerman2010information,bakshy11,goel12}, email forwarding \cite{WU2004327} and recommendation chains \cite{DynamicsOfViral}.

The majority of works focus on macro prediction tasks such as predicting the depth of a cascade, its size, or estimating other macro properties such as the transmission rate. Micro-level prediction tasks, such as whether a tweet $t$ will be retweeted or not, received considerably less attention. Even when microscopic properties were computed, like the probability that user $u$ will transmit information to $v$, it was used to derive macroscopic properties such as cascade size, and not for prediction in the standard machine-learning setting (train-test on a specific database of events). In \cite{galuba2010outtweeting}, two model-based algorithms for the task of url retweet prediction were proposed, based on the At-Least-One and Linear-Threshold schemes, and tested for a set of 100 popular URLs, whether they will be retweeted in the following time interval or not. In \cite{petrovic2011rt}, machine learning was used for the task of predicting for a stream of tweets which will get retweeted. 

We study a different setting, which is user-centric and not post-centric. Given a user $u \in V$, $V$ is some fixed set of users, and a time interval $[t_0,t_0 + \Delta T]$, the task is to predict whether user $u$ is going to react (retweet, reply, or mention) to an existing post by some user $v \in V$ (which $u$ doesn't necessarily follow). The input to the prediction algorithm is a snapshot of the activity of all users in $V$ at time $[t_0-\Delta T,t_0]$, namely a binary vector indicating whether each user was active or not. The user-centric task coincides with the post-centric task if all users react to at most one post. This is indeed the case for most users in our datasets, as Figure \ref{fig:act_hist} demonstrates.

In this work we address some fundamental questions regarding the nature in which information propagates in OSNs: (1) Can we perform the prediction task without any other features besides the time series of the users' postings (in particular, no linguistic features) (2) Is there an algorithm which performs this task, but is model-agnostic? namely, it does not rely on epidemiological models, and (3) Does the algorithm need to know the social links (which user follows/mentions which user)? Or can the algorithm implicitly learn this information from the snapshot it is given as input?

{\bf Our Contribution.}
We answer these questions, providing new insights into the way information propagates in online social media platforms. We design several prediction algorithms, each using a different approach, including a neural network (NN) based algorithm, $TWCRN$, which does not use the adjacency matrix of $V$. We evaluate our approach on four datasets, spanning a total of 77M tweets, which we collected from Twitter in 2020, following four different major events in that year. Our algorithm  $TWCRN$ performed the prediction task with an average F1 score of 0.86 over all datasets, which were highly imbalanced (on average, only 10\% of users in a time slice of width $\Delta T=12h$ reacted to some post).

We were able to verify that indeed $TWCRN$ implicitly learns information about the social links using a permutation test, where we randomly permuted the input to the NN at train time, contaminating social links information. The algorithm performed very poorly in this setting. 

We further designed a very intuitive algorithm, which for every user $u \in V$ learns the conditional probability that $u$ will re-share something in this time interval given $u$'s activity in the previous time interval (ignoring information about all other users). Intuitively, we are learning the ``trends" of $u$, is he a lazy-twitter (taking breaks between tweets) or a chain-tweeter. The performance of this algorithm was extremely competitive to $TWCRN$ (with respect to simplicity and running time), with an average  F1-score of 0.78. 

Thus we arrive at {\bf two new insights}. The first is that a relatively accurate prediction at the user's level can be obtained by simply learning the user's habits, regardless of the social links. The second is that a higher accuracy can be achieved by integrating information about social links, but, this information need not be served explicitly, but rather the algorithm can learn the relevant links by itself.

From a technical point of view, we are able to train a rich algorithm with many parameters, $TWCRN$, on a large set of users, $10^4$. Previous work either trained a simple model (perceptron) on a large set \cite{petrovic2011rt}, or a rich algorithm (Linear Threshold with many parameters) but trained on a small dataset (of size 100). To overcome the computational challenge we wrap the NN with an encoder-decoder pair, to reduce dimensionality. 

As far as we know, we are the first to use neural networks to perform such a prediction task, with such a success rate. The F1-score achieved in \cite{petrovic2011rt}, using many more features, and the passive-aggressive algorithm of \cite{PA} was 0.46, albeit for a somewhat different problem and setting. Neural networks were used very recently for the first time to predict macro properties of cascades (cascade structure), \cite{CascadesLSTM,li2017deepcas}, or to predict the influence of a given user based on activity features \cite{qiu2018deepinf}.

Our final contribution is to make all datasets openly available to the community, along with the Python code that we wrote for all algorithms.

\section{Related Work}
Users' influence in OSNs and the diffusion processes that stand behind  information propagation in them  were extensively studied in the past years. The works we survey here are by no means a concise review of the huge body of work available on this topic.

One line of work tries to maximize the size of the cascade by picking a good set of influential users that will be the starting seed. In \cite{matsubara2012rise}, the authors tried to predict users' influence by construction of differential equations extended from the Susceptible-Infected model. In \cite{tang2009relational}, the authors incorporated user-specific topic distribution and network structure, and in \cite{kempe2003maximizing, chen2010scalable} by means of a greedy algorithm. In \cite{feng2014influence}, the authors introduce the idea of novelty decay into the Independent Cascade model from \cite{albert2002statistical}. 

Another line of work is focused on understanding information diffusion as they unfold, without manipulating the seed. These works typically suggest a generative model that underpins the information cascade and then check how the model fits either simulation or real-world datasets.
Examples include \cite{zhao2015seismic, mishra2016feature} where a self-exciting point process is used to develop a statistical model to predict cascade size; cascades in heterogeneous scale-free networks were studied in \cite{moreno2002epidemic}  under the SIS and SIR model;
a cascade generating function was developed in \cite{Lerman11digg} to compute macroscopic properties of cascades, such as their size, spread, diameter, number of paths, and average path length. In \cite{li2017deepcas} an end-to-end deep learning approach was designed to predict the future size of cascades, without using any hand-crafted features or generative model assumptions.

The third line of work does not deal with prediction but rather with descriptive analysis, providing insights to the distribution of cascade size, depth, width and adoption rate (as time progresses) \cite{cha2009measurement,lerman2010information,ver2011stops,goel2016structural}.

Fewer works tried to predict microscopic post-level or user-level events. In \cite{galuba2010outtweeting}, the task was to predict whether a popular URL that was retweeted several times, will be retweeted in the following time window. The method was again model-based, upgrading the Linear Threshold (LT) and At-Least-One (ALO) models with additional features to achieve better prediction (F1 score of nearly 0.7). In \cite{petrovic2011rt} machine learning was used to predict retweets of an original tweet, albeit with a rather low F1 score (0.46). Our work continues this line of work but differs in two main aspects: we do not have any model-based assumptions as in \cite{galuba2010outtweeting}, and therefore we don't use any hand-crafted features (like the virality of the topic, number of followers, etc.), that were used also in \cite{petrovic2011rt}. Furthermore, our approach does not require an explicit description of the social links.


\section{Methodology} \label{methodology}
Previous studies on user influence in social networks in general, and Twitter specifically \cite{galuba2010outtweeting, petrovic2011rt}, point out that the main predictor of users' activity is the users he follows (i.e., his friends).

\begin{figure}
    \centering
    \includegraphics[width=0.5\textwidth]{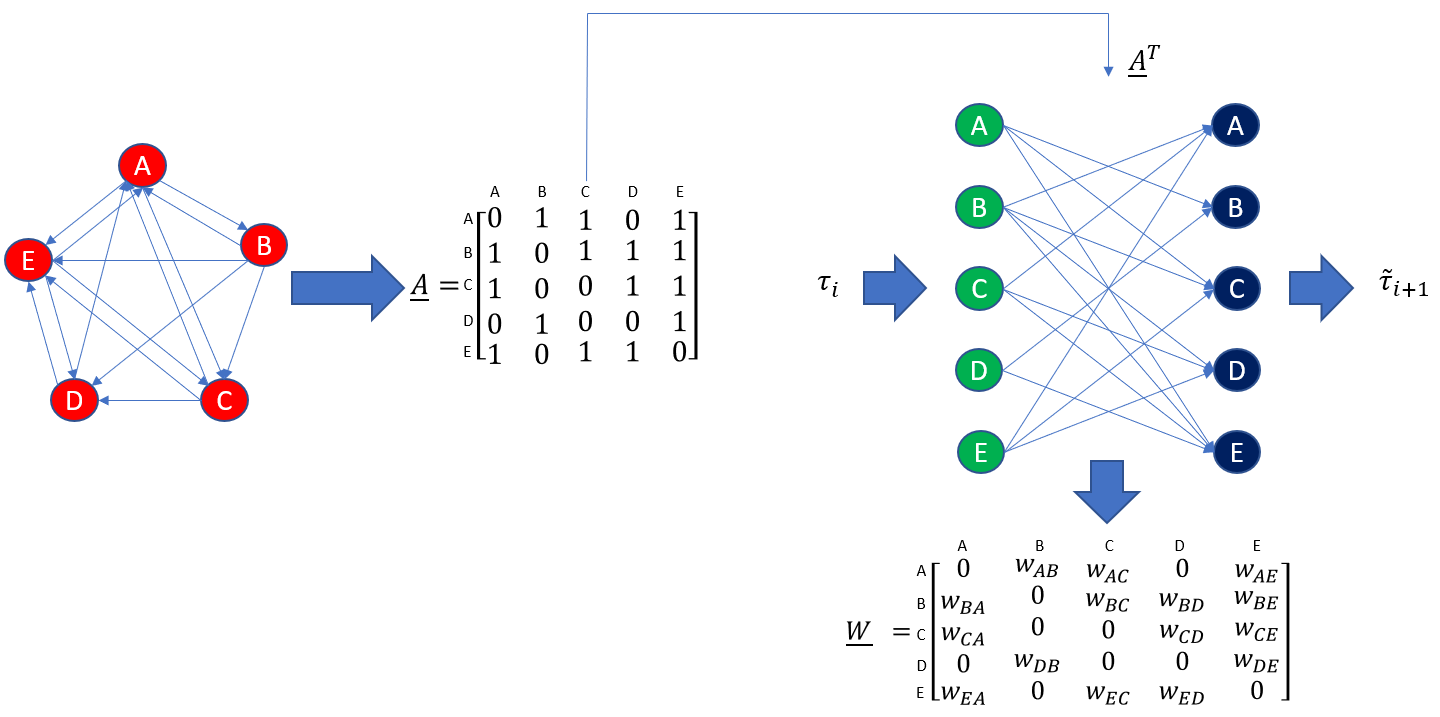}
    \caption{TWeet Mask Network ($TWMN$) schematic representation. The vector $\tau_i$ is the activity vector in the time window $[t_i - \Delta T, t_i]$, and $\tilde{\tau}_{i+1}$ is the predicted activity in time window $[t_i, t_i+ \Delta T]$.}
    \label{fig:tw_mask_net}
\end{figure}

We consider three models where the information about the user's  neighbors is taken into account to various degrees. In the first model, \emph{Tweet Prior Network} ($TWPN$) we completely ignore the user's social links and only consider its past activity, see Figure \ref{fig:tw_prior_net}. Our second model, the \emph{Tweet Mask Network} ($TWMN$), is a neural network that takes into account the user's social links and ignores non-adjacent users, Figure \ref{fig:tw_mask_net}. Finally, we train a fully-connected convolutional model, Tweet Convolutional Residual Network ($TWCRN$),  that considers all the activity of all the users in the network and not only the social links, Figure \ref{fig:tw_conv_res_net}.

The input to each NN is the same, a vector $\tau_i \in \{0,1\}^{|V|}$, which for every user $u \in V$ specifies whether he was active at time interval $[t_i-\Delta T,t_i]$ or not. The output vector, $\tilde{\tau}_{i+1} \in \R^{|V|}$ is the prediction for every user whether he is going to react to some post at time interval $[t_i,t_i+\Delta T]$ or not. 

\subsection{Models definition}
\label{models}
\begin{figure}
    \centering
    \includegraphics[width=0.30\textwidth]{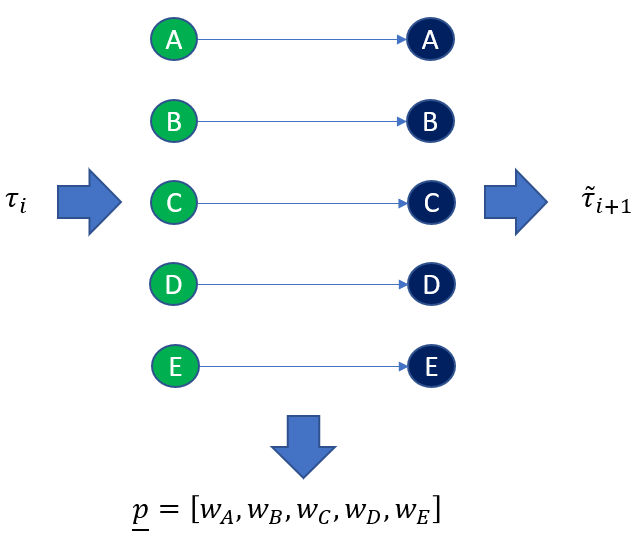}
    \caption{TWeet Prior Network ($TWPN$) schematic representation. The vector $\tau_i$ is the activity vector in the time window $[t_i - \Delta T, t_i]$, and $\tilde{\tau}_{i+1}$ is the predicted activity in time window $[t_i, t_i+ \Delta T]$.}
    \label{fig:tw_prior_net}
\end{figure}

\paragraph{Tweet Prior Network ($TWPN$)}
\label{tw_prior_net}
The architecture of this model is just two layers, input and output, as shown in Figure~\ref{fig:tw_prior_net}. All output nodes in the network use the $\tanh(w_u x_u)$ activation function, where $w_u$ is the weight of user's $u$ single edge, and $x_u$, the input, is the entry in $\tau_i$ corresponding to $u$.

\paragraph{MLE}
\label{mle_paragraph}
This model is a simpler variant of the $TWPN$ model where for every user $u \in V$,  the conditional probabilities are learned:
\begin{equation}
    p^{a,b}_{u}=Pr\left[I_u^{0} = a \, \big|\, I_u^{-1}= b\right],
    \label{eqn:mle}
\end{equation}

where, $I_u^{0}$ is the activity indicator of user $u$ at the predicted interval,  $I_u^{-1}$ in the activity indicator in the preceding time interval, and $a,b \in \{0,1\}$. The probabilities  $p^{a,b}_{u}$ are estimated solely on the data of user $u$ in the preceding time intervals, using the Maximum Likelihood Estimator, MLE (counting and averaging).

\paragraph{Tweet Mask Network ($TWMN$)}
This neural network extends $TWPN$ by taking into account the information at the user's neighbors as well. The architecture is a  fully connected 2-layer Feed-Forward Neural Network (FFNN) \cite{fine2006feedforward} (i.e., input layer fully connected to the output layer). Again, all output nodes use the $\tanh(\cdot)$ activation function. While the network is fully connected, for each output node $u$ we mask (hence the name) the non-adjacent input nodes.

For each user $u$, the output is given by $\tanh( \sum_{v \in N(u)} w_{v,u} x_v)$, where $x_v$ is the entry in $\tau_i$ corresponding to $v$. The weights $w_{v,u}$ between the user $u$ and the ones he follows $v \in N(u)$ can be interpreted as the influences, which the NN learns by itself. 

\begin{figure*}
    \centering
    \includegraphics[width=0.7\textwidth]{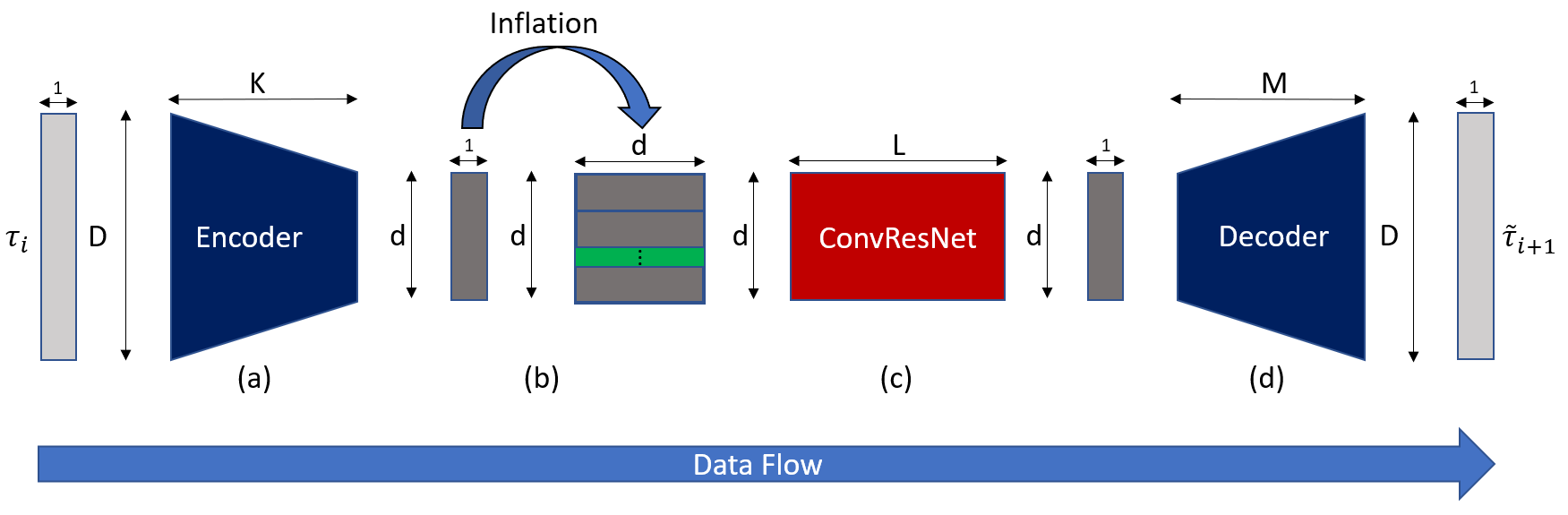}
    \caption{TW ConvResNet ($TWCRN$) schematic representation, where $\tau_i$ is the activity vector in the time window $[t_i - \Delta T, t_i]$, $\tilde{\tau}_{i+1}$ is the predicted activity in time window $[t_i, t_i+ \Delta T]$. \textbf{(a)}  the Encoder part of the Auto Encoder (AE) network with $K$ layers  \cite{hinton2006reducing}, which is responsible for compressing the original activity vector. \textbf{(b)} the inflation step, which receives a 1D vector and transforms it into a 2D matrix by copying it row-wise. \textbf{(c)}  the ConvResNet with $L=18$ or $L=34$ layers, as described in \cite{he2016deep}. \textbf{(d)} the Decoder network, which constitutes the second part of the AE network, with M layers, which is responsible for the final de-compression of the prediction vector.}
    \label{fig:tw_conv_res_net}
\end{figure*}

\paragraph{Tweet Convolutional Residual Network ($TWCRN$)}
This model takes into account all possible connections between users, no restricted to the follower (or mention) social links.
The model consists of four parts as shown in Figure~\ref{fig:tw_conv_res_net}, as follows: 

\vspace{1mm}

\noindent{\textbf{Encoder}} - A standard encoder, as described in \cite{hinton2006reducing}, which maps an activity vector of dimension $D$ (in our case $D=10^4$) to dimension $100 = d << D$. Each layer down samples the dimension of the input by a factor of 2 until the output size is $d=100$. The point of this layer is to reduce the computational effort of training the main part, ConvResNet, by reducing dimensionality.

\vspace{1mm}

\noindent{\textbf{Inflater}} - a simple procedure  which receives a vector of size $d$, and transforms it into a $d \times d$ matrix by copying it row-wise. This step is necessary because ConvResNet \cite{he2016deep} is used for images.

\vspace{1mm}

\noindent{\textbf{ConvResNet}} - We study separately two architectures, $\times 18$ and $\times 34$, presented in \cite{he2016deep} (the number represents the number of layers in the neural network)  which were developed for image processing tasks. This is the core of the prediction mechanism. The ConvResNet architecture enables extremely deep neural networks (more than 150 layers deep) to be successfully trained by a standard stochastic gradient descent with backpropagation, overcoming the problem of performance degradation present in very deep neural networks.

\vspace{1mm}

\noindent{\textbf{Decoder}} - Following \cite{hinton2006reducing}, the decoder maps the compressed vector (of dimension $d=100$) to its original dimension $D$. It does that by up-sampling the input vector by factor of 2, until the output size reaches the original input dimension $D$.

\subsection{Training and testing the models}\label{sec:train_test}
A cascade progresses in time, as well as in space (the nodes). In this paper, we focus on the time-aspect, and we partition the timeline along which the data was collected into equal-sized slices of $\Delta T=12$ hours each. For each pair of consecutive time periods $i,i+1$ we define two vectors $\tau_i,\tau_{i+1}$ as mentioned already above: $\tau_i$ is an indicator of  all activity types and $\tau_{i+1}$ only of reaction activity. 

The prediction task is, given the vector $\tau_i$, predict the reaction vector $\tau_{i+1}$. 

We train each model in the same manner, taking time period $i=1,\ldots,K$ as the train, and testing on time periods $i=K+1, \ldots N$,  where $N$ is the total number of time slices, and $K = \lfloor 0.9 \cdot N \rfloor$. We further split the train into train and validation, resulting in a 70-20-10 data split rule. To avoid over-fitting, we stopped the training if the loss on the validation set did not improve for 50 epochs.  

We used the ADAM optimizer \cite{kingma2014adam} with default settings for all three networks, $TWPN$, $TWMN$, and $TWCRN$ (the components of the $TWCRN$ model were trained separately). We chose the Mean Squared Error (MSE) for the loss function. To enrich the gradient flow in the training step, instead of binary vectors $\tau_i$ we used a technique in which each 1 was replaced with a sample from a normally distributed random variable $N(1, 0.01)$, and 0 with a sample from $N(-1, 0.01)$.

To find the conditional probabilities of the MLE model, $p^{a,b}_u$ in Eq.~\eqref{eqn:mle}, we counted for user $u$, the fraction of times that each combination $a,b$ appeared in the pairs $\tau_i,\tau_{i+1}$, for $i=1,\ldots,K$.

At the test step, we fed each model with $\tau_i$, for $i=K+1, \ldots N$, to receive the predicted vector $\tilde{\tau}_{i+1}$. Each entry in $\tilde{\tau}_{i+1}$ (either the output of tanh or of the decoder, depending on the model) is set to 1, if positive, and 0 if negative. The MLE probabilities were rounded to 0 or 1.
This vector is then compared with the real $\tau_{i+1}$ to compute precision, recall, and F1 score.


\section{Data}
\label{data}
We collected four datasets from Twitter, using the \emph{Tweepy} API. Each data set concerns a different major event at that time: \emph{Volcano} (the eruption of the Taal volcano in the Philippines on January 12, 2020), \emph{Kobe} (the death of Kobe Bryant in a helicopter crash on January 26, 2020), \emph{Princess} (the stepping down of the Duke and Duchess of the British royal family in January 2020) and \emph{Beirut} (the explosion in Beirut on August 4, 2020). Table \ref{tab:data} describes the keywords that were used and the size of each dataset, and Table \ref{tab:post_types} shows the breakdown according to different types of posts.

\begin{table*}
    \centering
    \begin{tabular}{l|l|l}
    \textbf{Name} & \textbf{Keyword} & \textbf{Posts}\\
    \hline
    Volcano & taal, volcano, philippines & 17,994,283\\
    Kobe & kobe, bryant & 24,562,000\\
    Princess & meghan, markle, harry, prince, princess & 22,353,429\\
    Beirut & beirut, lebanon, explosion & 14,106,557\\
    Unfiltered & - & 56,338,915\\
    \end{tabular}
    \caption{Description of the five datasets that we collected. The first four were collected during different "viral" events, which took place during the year 2020, and were filtered to include only tweets which were related to those events. The fifth dataset is an unfiltered tweet stream which was recorded during the 8/2021.}
    \label{tab:data}
\end{table*}

There are several possible ways of posting and interacting on Twitter. The simplest post is a \emph{tweet}, i.e., a post that the user writes himself and is not related to any other post. Tweets appear on the home timeline of the sender, and on the home timelines of all his followers, but Twitter's time-line algorithm keeps changing frequently. Features like ``in case you forgot" break the synchronous time-line structure, and suggestions of tweets from users one does not follow break the social link structure. 

There are three possible reactions to a tweet: \emph{mention} is a post that contains the "\emph{@username}" syntax in the body of the text, \emph{reply} is a direct response to another tweet and a \emph{re-tweet}, which is a propagation of someone else's tweet (re-tweet can be done also in an automatic manner using the 1-click option). All three reactions will be visible on the home timeline of the reacting user and on the timeline feed of all his followers. 

The next key component in the data pre-processing was to recover the social links between the users. We can infer two types of graphs. The first is the {\em mention graph}, in which a link exists between user $u$ and $v$ if $u$ reacted to $v$'s post (but $u$ need not follow $v$). Another graph, the {\em followers graph}, is the one recovered using Twitter's API where a link between $u$ and $v$ exists if user $u$ follows $v$. Note that there need not be an inclusion relationship between the two sets of links, but the way our dataset was collected (using the mention graph), entails that the followers graph is a subgraph of the mention graph.

We recovered both social networks, and as we shall see, the accuracy of the classification task differs according to which network we use. Let us note that querying Twitter's API is computationally more expensive as Twitter limits the number of requests. Therefore it is interesting to know what's the gain, if any, when using the more costly-obtained information of the Twitter followers graph.

\begin{table*}
    \centering
    \begin{tabular}{l|l|l|l|l|l}
    \textbf{Name} & \textbf{PO} & \textbf{TW} & \textbf{RT} & \textbf{MT}\\
    \hline
        Volcano & 17,994,283 & 3,400,415 & 11,810,708 & 2,783,160 \\
        Kobe & 24,562,000 & 2,962,387 & 20,768,520 & 831,093 \\
        Princess & 22,353,429 & 4,694,745 & 15,457,234 & 2,201,450 \\
        Beirut & 14,106,557 & 1,951,752 & 11,418,457 & 736,348 \\
        Unfiltered & 56,338,915 & 35,728,226 & 813,013 & 19,797,676 \\
    \end{tabular}
    \caption{Breakdown of the post types for each data set, where \textbf{PO} is a general post (i.e., tweet, re-tweet or mention), \textbf{TW} is the initial tweet, \textbf{RT} is a re-tweet using the "\emph{RT @username}" syntax, or by clicking the dedicated button and \textbf{MT} is a mention.}
    \label{tab:post_types}
\end{table*}

\begin{figure}
    \centering
    \includegraphics[width=0.5\textwidth]{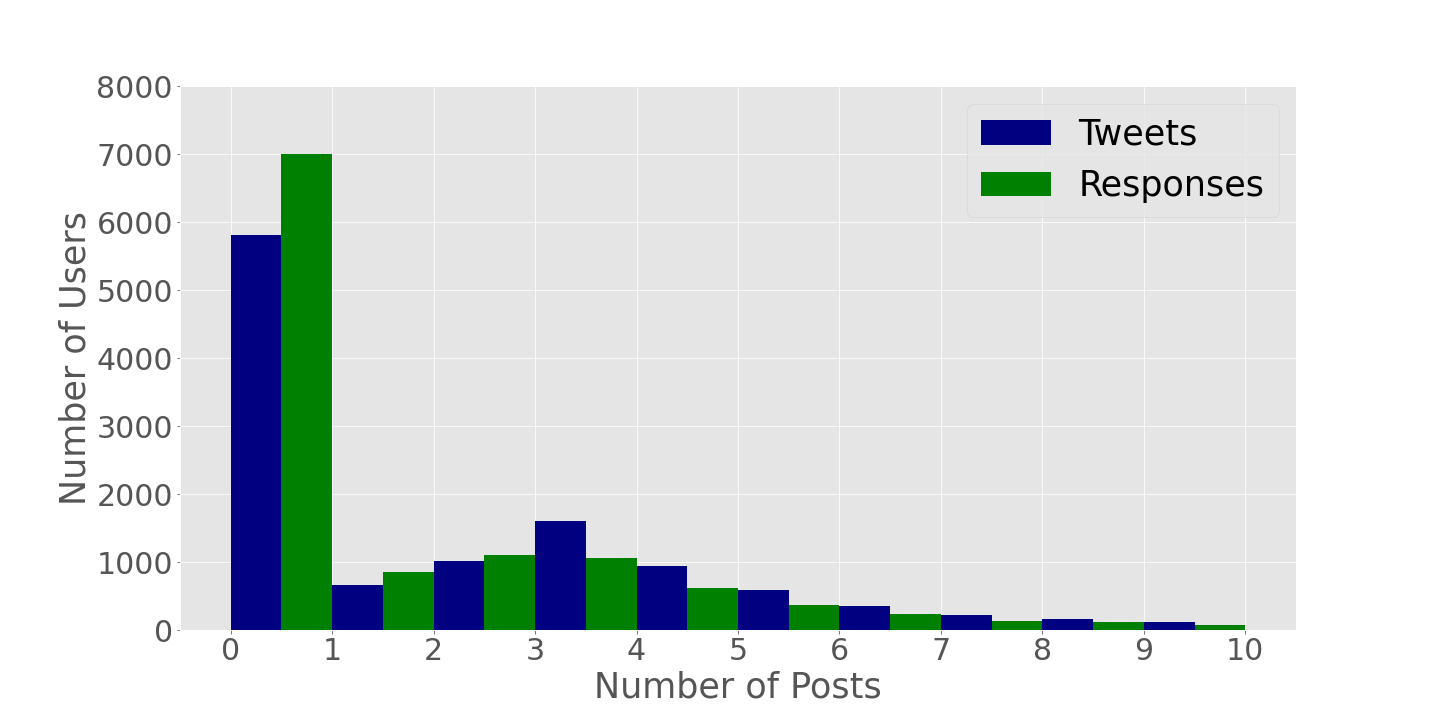}
    \includegraphics[width=0.5\textwidth]{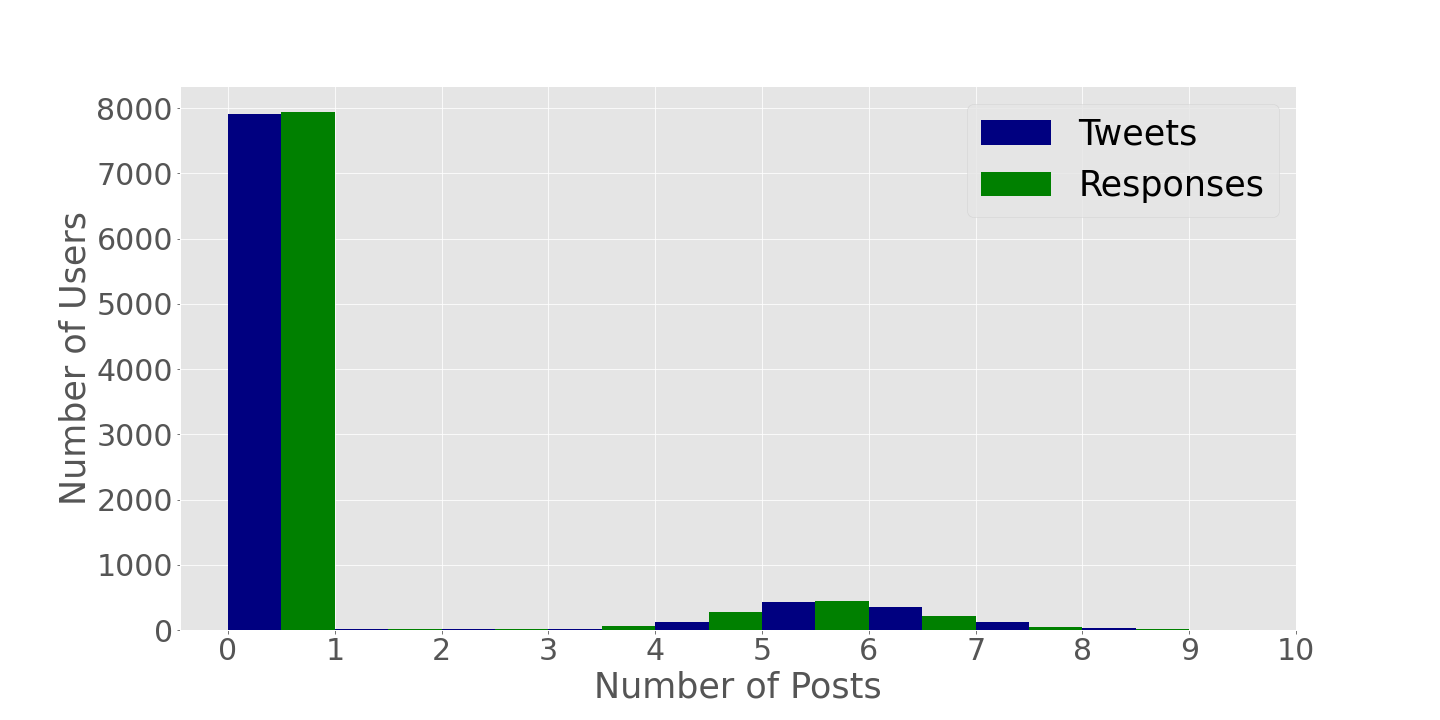}
    \caption{Histogram of the average activity of users in a single time slice of $\Delta T = 12h$ for the Princess dataset. Top: statistics for the mention graph dataset. Bottom: statistics for the followers graph dataset. Evidently, most users tweet or re-share only one post per time slice.}
    \label{fig:act_hist}
\end{figure}

As customary in many works, users that have low activity are filtered out from the dataset. We created two sets, $A$, the active users, users who post at least $a$ posts (either tweet or re-share of some sort), and $P$, the popular users, users that were reacted to at least $p$ times. We chose $a$ and $p$ so that the number of users in the dataset is roughly 10,000 (due to the constraint of memory resources). 

\paragraph{Broadcasticity.}
Another parameter that is of interest is the depth of the cascade. As noted, most cascades are shallow. Various measures were used to compute depth, including the Wiener index in \cite{goel2016structural}, the average distance between all pairs of nodes in a diffusion tree. We suggest a much simpler criterion, which does not require the laborious task of computing the diffusion tree. We call it the {\em broadcasticity} measure, and it measures how much the diffusion resembles a broadcast rather than a diffusion. It's computed using the Jaccard index of the two sets $A$ and $P$:

\begin{equation}
    B = 1 - J(A, P) = 1 - \frac{|A \cap P|}{|A \cup P|}
    \label{eqn:broadcastisity}
\end{equation}

High broadcasticity means that $A$ and $P$ are almost disjoint, namely users either tweet or react, but very few do both. This is exactly the case of a shallow 1-hop diffusion. If a diffusion on the other hand is deep and wide   (like an epidemic spreading in the population), then many users belong to both $A$ and $P$, and the broadcasticity is low. Note that a shallow diffusion with few long paths still attains large broadcastisity.

Table \ref{tab:data_broadcastisity} describes the various features of each dataset with respect to $A$, $P$ and the broadcastisity.
\begin{table*}
    \centering
    \begin{tabular}{l|l|l|l|l|l|l|l}
    \textbf{Name} & \textbf{$a$} & \textbf{$|A|$} & \textbf{$p$} & \textbf{$|P|$} & \textbf{$Posts$} & \textbf{\#slices}  & $B$ \\
    \hline
        Volcano & 103 & 3,189 & 207 & 7,757 & 2,710,915 & 108 & 0.97  \\
        Kobe & 167 & 4,674 & 167 & 4,835 & 1,380,543 & 87 & 0.98 \\
        Princess & 234  & 5,151 & 234 & 7,866 & 2,710,915 & 87 & 0.92 \\
        Beirut & 105 & 3,830 & 210 & 5,778 & 960,228 & 117 & 0.93 \\
        Unfiltered & 100 & 3,943 & 201 & 5,238 & 721,301 & 28 & 0.99 \\
    \end{tabular}
    \caption{Datasets are filtered to remove low-activity users. The set $A$ of active users that (re-) share at least $a$ posts (over the entire time period), and popular users $P$ whose tweets were re-shared at least $p$ times. The number of time slices $\tau_i$ is given in column 7, and the broadcastiscity in the last column.}
    \label{tab:data_broadcastisity}
\end{table*}

\begin{table*}
    \centering
    \begin{tabular}{l|c|c|c|c|c|c}
    \textbf{Name} & \multicolumn{3}{c|}{\textbf{Network Crawl}} & \multicolumn{3}{c}{\textbf{Tweet Log}}\\
    \hline
        & N & $E_c[L]$ & $E[L]$ & N & $E_c[L]$ & $E[L]$ \\
        \cline{2-7}
        Volcano & 8,990 & 3.953 & \color{green}{2.943} & 10,440 & 4.019 & \color{red}{4.648} \\
        Kobe & 8,339 & 3.921 & \color{green}{2.368} & 9,225 & 3.965 & \color{green}{2.593} \\
        Princess & 9,793 & 3.991 & \color{green}{2.869} & 11,891 & 4.075 & \color{green}{3.579} \\
        Beirut & 7,943 & 3.9 & \color{green}{2.498} & 8,876 & 3.948 & \color{green}{3.103} \\
        Unfiltered & 7,208 & 3.86 & \color{red}{6.23} & 7,363 & 3.87 & \color{red}{6.23} \\
    \end{tabular}
    \caption{Mean shortest path ($E[L]$) between any two randomly chosen nodes in the social networks. The $E_c[L] = log(N)$, where $N$ is the number of \textbf{valid} users in the social networks, and $E_c[L]$ is the critical mean shortest path (i.e., the network should satisfy the condition $E[L] \leq E_c[L]$, to be considered a small world (\textbf{SW}) network). \color{green}{Green} - the criterion is satisfied, \color{red}{red} - the criterion is not satisfied.}
    \label{tab:mean_shortest_path}
\end{table*}


\section{The Experiment}
Recall the classification task described in Section \ref{sec:train_test}. Given a vector $\tau_{i} \in \{0,1\}^{|V|}$, in which the $j^{th}$ entry is 1 if user $j$ tweeted in time window $i$, our goal is to predict the vector $\tau_{i+1}$, in which the $j^{th}$ entry is 1 if user $j$ reacted to some post in time interval $i+1$.
\subsection{Other baselines}\label{sec:baseline}
We benchmarked our model against four other popular models. The first two are  random guess models $RND_{p=0.5}$ and $RND_{p=\pi}$ which predict $\tau_{i+1}[j]$ using a fair random coin flip, or the fraction of users that were active in the previous time window, $\pi$. 
The two last models are variations of the SI model, At-Least-One ($ALO$) and Linear-Threshold ($LT$). We follow the $LT$ and $ALO$ models that were  developed in  \cite{galuba2010outtweeting} as we found them the most general. The intuition of both is that if the neighbors of the user  were active in the preceding time window, it may stir the user's decision to become active himself in the next time window. Each algorithm takes a different try at estimating the probability that user $u$ will become active in the next time window; \cite{galuba2010outtweeting} use the following expression for $LT$:
\begin{equation}
    p_{u}^{LT} = A\Big(\alpha_{v_1,u},\ldots,\alpha_{v_k,u}, \beta_{u}, \gamma\Big)T\Big(\mu_{u}, \sigma_{u}^2, t_{u}^{post}\Big)
    \label{eq:p_LT}
\end{equation}
The $\alpha_{v_i,u} \in [0, 1]$ are the influence of a neighbor of $u$, $v_i$, on $u$. The $\beta_{u} \in [0, 1]$ is a prior probability of user $u$ to become active, $\gamma \in [0, 1]$ is the virality of the topic that is being discussed. $A(\cdot)$ is an \textbf{a-}temporal component, i.e., the probability that the user will respond to some post due to the influence that was exerted on him from his social circle. Concretely, this component is given by

 \begin{equation*}
        A(\cdot) = \sigma_{a, b}\bigg(\gamma\Big(\beta_{u} + \sum_{v:v \rightarrow u}\gamma\alpha_{v,u}p_{u}\Big)\bigg),
    \end{equation*}
    
where $\sigma_{a, b}(x) = \frac{1}{1+e^{-a(b-x)}}$ is a sigmoid function.

The second component of  Eq.~\ref{eq:p_LT}, $T(\cdot)$, is a temporal component of the predicted activity probability, and it represents an empirical observation regarding the time that passes from a moment when any user in the social network initiates a new post until the first of his followers responds to it ($\mu_{u}$ and $\sigma_{u}^2$ are the mean and SD of the latter). This component is unique for each social network, and is given by

\begin{equation*}
    T\Big(\mu_{u}, \sigma_{u}^2, t_{u}^{post}\Big) = \frac{1}{2}erfc\Big(-\frac{ln(t_{u_i}^{post})-\mu_{u}}{\sigma_{u}\sqrt{2}}\Big)
    \label{eqn:temp}
\end{equation*}

The parameter $t_{u_i}^{post}$ represents the time of post's initiation relative to the start of the time window, and $erfc(x) = 1 - \frac{2}{\sqrt{\pi}}\int_{0}^{x} e^{-t^2}dt$, is a complementary Gauss error function.\\
For $ALO$ the following is used in \cite{galuba2010outtweeting}:
    \begin{equation}
       p_u^{ALO}= 1 - \Big(1-\gamma\beta_{u}\Big)\prod_{v:v \rightarrow u}\Big(1-\gamma\alpha_{v,u}p_{u}\Big)
        \label{eq:p_ALO}
    \end{equation}

In both cases, we followed the rule in \cite{galuba2010outtweeting}, where a user is predicted to react if $p_u>0.5$

\subsection{Training the baseline models}
The base-line models were trained similarly to our models, using 90\% of the time intervals for train and 10\% for test.
Table \ref{fig:act_hist} shows the total number of time intervals per dataset.

$RND_{p=0.5}$ and $RND_{p=\pi}$ have no tunable parameters.
The parameters of $ALO$ and $LT$ were set as follows. Optimization was performed with Python's \texttt{Hyperopt} library, with the F1 score as the function to maximize. This library uses the
Tree of Parzen Estimators hyper-parameter optimization algorithm from \cite{bergstra2011algorithms}. The parameter $\gamma$, topic virality, was set to 1, since all tweets come from the same topic. Also the temporal part $T(\cdot)$ was set to 1, since we are not interested in the event that a tweet $t$ is re-shared sometime in the future, but whether user $u$ re-shared some content in a very limited time interval. Due to the computational constrains, the optimization was performed only on the prior vector of the $p_u$'s. The dataset in \cite{galuba2010outtweeting} included 100 URLs for which retweeting was predicted, ours include 10,000 events, and the number of $\alpha_{u,v}$'s is even larger. Therefore, the influence matrix $\alpha_{u,v}$ was supplied from the output of the $TWMN$ model, where the weight of the edge in the NN that connects $v$ and $u$ may be thought of as the influence of $v$ on $u$.

\begin{table*}
    \centering
    \begin{tabular}{l|l|l|l|l}
    \textbf{Model} & \textbf{CPU} & \textbf{RAM} & \textbf{Accelerator} & \textbf{Run Time}\\
    \hline
    $RND_{p=0.5}$ &  $CPU_{low}$ & 32 GB & None & $31.75\pm1.48[s]$\\
    $RND_{p=prop}$ &  $CPU_{low}$ & 32 GB & None & $30.00\pm1.22[s]$\\
    $MLE$ &  $CPU_{low}$ & 32 GB & None & $28.38\pm1.41[s]$\\
    \hline
    $ALO$ &  $CPU_{high}$ & 16 GB & None & $1.45\pm0.38[H]$\\
    $LT$ &  $CPU_{high}$ & 16 GB & None & $1.29\pm0.31[H]$\\
    \hline
    $TWMN_{SHUF}$ &  $CPU_{high}$ & 16 GB & GPU & $1.18\pm0.75[H]$\\
    $TWMN_{all 1}$ &  $CPU_{high}$ & 16 GB & GPU & $54.35\pm22[m]$\\
    $TWMN$ &  $CPU_{high}$ & 16 GB & GPU & $1.12\pm72[H]$\\
    \hline
    $TWPN$ &  $CPU_{high}$ & 16 GB & GPU & $1.29\pm54[H]$\\
    \hline
    $TWCRN_{x18}^{SHUF}$ &  $CPU_{high}$ & 16 GB & GPU & $1.1\pm0.28[H]$\\
    $TWCRN_{x18}$ &  $CPU_{high}$ & 16 GB & GPU & $54\pm11.5[m]$\\
    \hline
    $TWCRN_{x34}^{SHUF}$ & $CPU_{high}$ & 16 GB & GPU & $1.23\pm0.32[H]$\\
    $TWCRN_{x34}$ & $CPU_{high}$ & 16 GB & GPU & $1 \pm 0.23[H]$\\
 
    \end{tabular}
    \caption{Hardware specs with corresponding mean run times, where $CPU_{low}$ is Intel i7-8550U @ 1.80GHz, $CPU_{high}$ is Intel Xeon @ 2.00GHz and GPU is Tesla P100-PCIE-16GB, [H], [m] and [s] stands for "Hours", "minutes" and "seconds" respectively. }
    \label{tab:hardware_runtimes}
\end{table*}

\begin{landscape}
\begin{table*}
  \centering
  \resizebox{\columnwidth}{!}{\begin{tabular}{|l||*{15}{c|}}
    \hline
    \textbf{Metric} & 
    \textbf{P} & \textbf{$F_1$} & \textbf{R} & 
    \textbf{P} & \textbf{$F_1$} & \textbf{R} & 
    \textbf{P} & \textbf{$F_1$} & \textbf{R} & 
    \textbf{P} & \textbf{$F_1$} & \textbf{R} & 
    \textbf{P} & \textbf{$F_1$} & \textbf{R} \\
    \hline\hline
    $RND_{p=0.5}$ & 
    \mathval{0.19} & \mathval{0.28} & \mathval{0.5} & 
    \mathval{0.29} & \mathval{0.34} & \mathval{0.5} & 
    \mathval{0.34} & \mathval{0.4} & \mathval{0.5} & 
    \mathval{0.17} & \mathval{0.23} & \mathval{0.5} & 
    \mathval{0.33} & \mathval{0.4} & \mathval{0.5}\\ 
    \hline
    $RND_{p=\pi}$ & 
    \mathval{0.19} & \mathval{0.21} & \mathval{0.24} & 
    \mathval{0.29} & \mathval{0.29} & \mathval{0.34} & 
    \mathval{0.34} & \mathval{0.38} & \mathval{0.43} & 
    \mathval{0.17} & \mathval{0.18} & \mathval{0.22} & 
    \stdcell{0.63}{0.02} & \stdcell{0.64}{0.01} & \stdcell{0.64}{0.01} \\ 
    \hline
    $MLE$ & 
    \mathval{0.42} & \mathval{0.42} & \mathval{0.46} & 
    \stdcell{0.31}{0.02} & \stdcell{0.38}{0.02} & $0.56$ & 
    \stdcell{0.58}{0.02} & \mathval{0.6} & \mathval{0.68} & 
    \mathval{0.22} & \mathval{0.29} & \mathval{0.44} & 
    \stdcell{0.63}{0.01} & \stdcell{0.63}{0.01} & \stdcell{0.64}{0.01} \\ 
    \hline\hline
    $ALO$ & 
    \stdcell{0.18}{0.02} & \stdcell{0.26}{0.03} & $0.52$ & 
    \stdcell{0.18}{0.06} & \stdcell{0.26}{0.07} & $0.51$ & 
    \stdcell{0.31}{0.03} & \stdcell{0.38}{0.04} & $0.53$ & 
    \stdcell{0.09}{0.02} & \stdcell{0.15}{0.03} & \stdcell{0.51}{0.02} & 
    \mathval{0.4} & \mathval{0.54} & \mathval{0.83}\\ 
    \hline
    $LT$ & 
    \stdcell{0.18}{0.03} & \stdcell{0.26}{0.04} & \mathval{0.52} & 
    \stdcell{0.17}{0.04} & \stdcell{0.24}{0.05} & \mathval{0.5} & 
    \stdcell{0.31}{0.05} & \stdcell{0.37}{0.06} & \stdcell{0.53}{0.01} & 
    \stdcell{0.11}{0.03} & \stdcell{0.17}{0.05} & \stdcell{0.53}{0.01} & 
    \mathval{0.17} & \mathval{0.17} & \mathval{0.17} \\ 
    \hline\hline
    $TWMN_{all 1}$ & 
    \mathval{0.49} & \mathval{0.45} & \mathval{0.49} & 
    \stdcell{0.31}{0.02} & \stdcell{0.38}{0.01} & \mathval{0.59} & 
    \stdcell{0.57}{0.01} & \stdcell{0.57}{0.01} & \mathval{0.66} & 
    \mathval{0.21} & \mathval{0.29} & \mathval{0.53} & 
    $0.39$ & \stdcell{0.46}{0.02} & \stdcell{0.56}{0.02} \\ 
    \hline
    $TWMN$ & 
    \mathval{0.47} & \mathval{0.48} & \best{0.52} & 
    \stdcell{0.32}{0.02} & \stdcell{0.44}{0.02} & \best{0.83} & 
    \stdcell{0.6}{0.03} & \stdcell{0.63}{0.03} & \stdcell{0.75}{0.01} & 
    \mathval{0.26} & \mathval{0.32} & \mathval{0.46} & 
    \mathval{0.64} & \mathval{0.64} & \mathval{0.65} \\ 
    \hline\hline
    $TWPN$ & 
    \mathval{0.51} & \mathval{0.49} & \mathval{0.5} & 
    \stdcell{0.33}{0.01} & \stdcell{0.45}{0.01} & $0.8$ & 
    \mathval{0.63} & \mathval{0.66} & \best{0.78} & 
    \best{0.327} & $0.36$ & $0.45$ & 
    \mathval{0.68} & \mathval{0.68} & \mathval{0.68} \\ 
    \hline\hline
    $TWCRN_{x18}^{SHUF}$ & 
    \mathval{0.23} & \mathval{0.2} & \mathval{0.18} & 
    \mathval{0.22} & \mathval{0.3} & \stdcell{0.48}{0.02} & 
    \mathval{0.41} & \mathval{0.42} & \stdcell{0.43}{0.01} & 
    \mathval{0.15} & \mathval{0.19} & \stdcell{0.25}{0.02} & 
    \mathval{0.59} & \mathval{0.58} & \mathval{0.58} \\ 
    \hline
    $TWCRN_{x18}$ & 
    \stdcell{0.54}{0.02} & \stdcell{0.51}{0.01} & $0.49$ & 
    \best{0.38} & \beststdcell{0.5}{0.02} & \stdcell{0.75}{0.07} & 
    \mathval{0.7} & \mathval{0.73} & \mathval{0.77} & 
    \stdcell{0.32}{0.02} & \beststdcell{0.42}{0.02} & \beststdcell{0.62}{0.02} & 
    \mathval{0.837} & \best{0.836} & \best{0.835} \\ 
    \hline\hline
    $TWCRN_{x34}^{SHUF}$ & 
    \mathval{0.23} & \mathval{0.2} & \mathval{0.18} & 
    \mathval{0.21} & \stdcell{0.28}{0.02} & \stdcell{0.45}{0.03} & 
    \mathval{0.41} & \mathval{0.42} & \mathval{0.44} & 
    \mathval{0.15} & \mathval{0.19} & \stdcell{0.25}{0.02} & 
    \stdcell{0.59}{0.01} & \stdcell{0.58}{0.01} & \stdcell{0.58}{0.01} \\ 
    \hline
    $TWCRN_{x34}$ & 
    \beststdcell{0.55}{0.01} & \best{0.52} & \stdcell{0.49}{0.01} & 
    \mathval{0.36} & \stdcell{0.49}{0.01} & \stdcell{0.79}{0.03} & 
    \best{0.71} & \best{0.74} & \stdcell{0.777}{0.01} & 
    \stdcell{0.32}{0.02} & \stdcell{0.419}{0.01} & \stdcell{0.61}{0.04} & 
    \beststdcell{0.838}{0.01} & \stdcell{0.835}{0.01} & \stdcell{0.832}{0.02} \\ 
    \hline\hline
  \end{tabular}}
  \label{tab:mt_graph_res}
  \caption{Summary of a 5-fold validation results on the test set for the datasets acquired through the mention graph technique, i.e., each cell is the average of five train-test executions ((most of the mean values were rounded up to the second decimal point, unless a higher precision was needed to decide on the best value, and standard deviations smaller then $0.01$ are omitted). The best values for each metric are represented by the \textcolor{green}{green} color.}
\end{table*}
\end{landscape}

\begin{landscape}
\begin{table*}
  \centering
  \resizebox{\columnwidth}{!}{\begin{tabular}{|l||*{15}{c|}}
    \hline
    \backslashbox{\textbf{Model}}{\textbf{Data Set}} & \multicolumn{3}{c|}{\textbf{Volcano}} & 
    \multicolumn{3}{c|}{\textbf{Kobe}} &
    \multicolumn{3}{c|}{\textbf{Princess}} &
    \multicolumn{3}{c|}{\textbf{Beirut}} &
    \multicolumn{3}{c|}{\textbf{Unfiltered}} \\
    \hline
    \textbf{Metric} & 
    \textbf{P} & \textbf{$F_1$} & \textbf{R} & 
    \textbf{P} & \textbf{$F_1$} & \textbf{R} & 
    \textbf{P} & \textbf{$F_1$} & \textbf{R} & 
    \textbf{P} & \textbf{$F_1$} & \textbf{R} & 
    \textbf{P} & \textbf{$F_1$} & \textbf{R} \\
    \hline\hline
    $RND_{p=0.5}$ & 
    \mathval{0.1} & \mathval{0.16} & \mathval{0.5} & 
    \mathval{0.1} & \mathval{0.17} & \mathval{0.5} & 
    \mathval{0.11} & \mathval{0.18} & \mathval{0.5} & 
    \mathval{0.09} & \mathval{0.15} & \mathval{0.5} & 
    \mathval{0.31} & \mathval{0.39} & \mathval{0.5} \\ 
    \hline
    $RND_{p=\pi}$ & 
    \mathval{0.1} & \mathval{0.1} & \mathval{0.1} & 
    \mathval{0.1} & \mathval{0.1} & \mathval{0.11} & 
    \mathval{0.11} & \mathval{0.12} & \mathval{0.12} & 
    \mathval{0.09} & \mathval{0.09} & \mathval{0.1} & 
    \mathval{0.32} & \mathval{0.4} & \mathval{0.55} \\ 
    \hline
    $MLE$ & 
    \mathval{0.8} & \mathval{0.81} & \mathval{0.86} & 
    \stdcell{0.73}{0.03} & \stdcell{0.78}{0.02} & \mathval{0.88} & 
    \mathval{0.88} & \mathval{0.89} & \mathval{0.97} & 
    \stdcell{0.59}{0.01} & \mathval{0.65} & \mathval{0.79} & 
    \mathval{0.63} & \mathval{0.64} & \mathval{0.64}\\ \hline\hline
    $ALO$ & 
    \mathval{0.09} & \stdcell{0.16}{0.01} & \mathval{0.51} & 
    \mathval{0.08} & \stdcell{0.14}{0.03} & \stdcell{0.5}{0.01} & 
    \stdcell{0.11}{0.1} & \stdcell{0.18}{0.02} & \stdcell{0.53}{0.02} & 
    \stdcell{0.07}{0.01} & \stdcell{0.12}{0.02} & \stdcell{0.54}{0.01} & 
    \mathval{0.32} & \mathval{0.47} & \mathval{0.87} \\ 
    \hline
    $LT$ & 
    \stdcell{0.1}{0.01} & \stdcell{0.17}{0.02} & \mathval{0.56} & 
    \stdcell{0.09}{0.02} & \stdcell{0.15}{0.03} & \mathval{0.56} & 
    \stdcell{0.12}{0.01} & \stdcell{0.19}{0.02} & \mathval{0.55} & 
    \stdcell{0.07}{0.01} & \stdcell{0.12}{0.02} & \stdcell{0.56}{0.01} & 
    \mathval{0.31} & \mathval{0.2} & \mathval{0.15} \\ 
    \hline\hline
    $TWMN_{all 1}$ & 
    \mathval{0.81} & \mathval{0.81} & \mathval{0.85} & 
    \stdcell{0.74}{0.02} & \stdcell{0.8}{0.01} & \mathval{0.93} & 
    \mathval{0.86} & \mathval{0.86} & \mathval{0.97} & 
    \stdcell{0.6}{0.02} & \stdcell{0.66}{0.17} & \mathval{0.79} & 
    \stdcell{0.37}{0.03} & \stdcell{0.44}{0.02} & \stdcell{0.55}{0.01} \\ 
    \hline
    $TWMN$ & 
    \mathval{0.8} & \mathval{0.86} & \mathval{0.97} & 
    \stdcell{0.74}{0.02} & \stdcell{0.83}{0.02} & \best{0.997} & 
    \mathval{0.86} & \mathval{0.87} & \mathval{0.99} & 
    \mathval{0.56} & \mathval{0.69} & \mathval{0.96} & 
    \mathval{0.47} & \mathval{0.54} & \mathval{0.62} \\ 
    \hline\hline
    $TWPN$ & 
    \mathval{0.8} & \mathval{0.86} & \mathval{0.97} & 
    \stdcell{0.73}{0.03} & \stdcell{0.82}{0.02} & \mathval{0.98} & 
    \mathval{0.86} & \mathval{0.87} & \mathval{0.99} & 
    \stdcell{0.58}{0.02} &  \stdcell{0.71}{0.02} & \mathval{0.96} & 
    \mathval{0.69} & \mathval{0.68} & \mathval{0.68} \\ 
    \hline\hline
    $TWCRN_{x18}^{SHUF}$ & 
    \mathval{0.09} & \mathval{0.01} & \mathval{0.1} & 
    \mathval{0.09} & \stdcell{0.1}{0.01} & \stdcell{0.12}{0.01} & 
    \mathval{0.1} & \mathval{0.1} & \mathval{0.1} & 
    \mathval{0.08} & \stdcell{0.09}{0.01} &  \stdcell{0.12}{0.01} & 
    \mathval{0.56} & \stdcell{0.55}{0.01} & \mathval{0.55} \\ 
    \hline
    $TWCRN_{x18}$ & 
    \best{0.892} & \mathval{0.927} & \mathval{0.966} & 
    \best{0.785} & \best{0.877} & \mathval{0.995} & 
    \best{0.975} & \mathval{0.9832} & \mathval{0.992} & \beststdcell{0.71}{0.01} & 
    \best{0.82} & \mathval{0.97} & \best{0.842} & \best{0.83} & \best{0.82} \\ 
    \hline\hline
    $TWCRN_{x34}^{SHUF}$ & 
    \mathval{0.11} & \mathval{0.11} & \mathval{0.12} & 
    \mathval{0.1} & \mathval{0.11} & \mathval{0.12} & 
    \mathval{0.11} & \mathval{0.11} & \mathval{0.11} & 
    \mathval{0.1} & \stdcell{0.1}{0.01} & \stdcell{0.12}{0.01} & 
    \mathval{0.56} & \mathval{0.56} & \mathval{0.551} \\ 
    \hline
    $TWCRN_{x34}$ &  
    \mathval{0.891} & \best{0.929} & \best{0.97} &  
    \mathval{0.78} & \mathval{0.876} & \mathval{0.996} &  
    \mathval{0.974} & \best{0.9833} & \best{0.993} &  
    \stdcell{0.68}{0.02} & \stdcell{0.8}{0.01} & \beststdcell{0.977}{0.01} & 
    \mathval{0.84} & \mathval{0.827} & \stdcell{0.81}{0.01} \\ 
    \hline\hline
  \end{tabular}}
  \label{tab:crwl_graph_res}
  \caption{Summary of results on the test set for the datasets acquired through the follower graph technique, i.e., each cell is the average of five train-test executions (most of the mean values were rounded up to the second decimal point, unless a higher precision was needed to decide on the best value, and standard deviations smaller then $0.01$ are omitted). The best values for each metric are represented by the \textcolor{green}{green} color.}
\end{table*}
\end{landscape}

\subsection{Results}
We turn to describe the results of evaluating both the algorithms described in Section \ref{methodology} and the baseline algorithms, Section \ref{sec:baseline}, on the data that we collected, Section \ref{data}.

Our raw dataset spans the mention graph. Namely, the social links are derived from users re-sharing posts of other users. We further removed low-activity users from the dataset, remaining with roughly $10^4$ users as described in Table \ref{tab:data_broadcastisity}. In our experiments, we ran the algorithms also on the followers graph dataset, which we obtained from the mention graph dataset by removing users which no one follows or they follow no one, by querying Twitter's API. 

\vspace{1mm}

To assess the impact of using the adjacency matrix $A$ on the performance of $TWMN$ we trained a second variant, $TWMN_{all-1}$, in which the masking is removed, namely the NN becomes fully connected (the same as replacing $A$ with the all one matrix). 

The convolutional network $TWCRN$ does not receive explicitly the social links. To understand whether it implicitly learns them or not, we performed the following permutation test, which resulted in a model we call $TWCRN_{SHUF}$. At train time, the input vectors $\tau_i$ that are fed into the model are randomly shuffled, anew for every $i$, while the labels of the result vectors $\tau_{i+1}$ were kept without change. 

Table \ref{tab:hardware_runtimes} describes the hardware and the running time it took to train each model. The Table \ref{tab:crwl_graph_res} describes the results for the mention graph dataset, while the Table \ref{tab:mt_graph_res}: describes the results for the followers graph dataset. Since the training of the algorithms contains random choices, we repeated each execution five times. The standard deviations in both tables are over these five executions. 

Note that our dataset is imbalanced. There are more zeros than ones in each $\tau_i$. The average density of $\tau_i$ is given by the  precision of the $RND_{p=0.5}$ algorithm.  

\vspace{1mm}

We summarize our main findings that we read from Tables Table \ref{tab:mt_graph_res} and \ref{tab:crwl_graph_res}:

\begin{itemize}
    \item Our first observation is that the results obtained from the followers graph dataset are significantly better for all four datasets, with F1 score nearly twice as good in three out of the four (for the leading method). This means that the signal embedded in the followers graph is stronger than the one embedded in the mention graph. Trying to understand why, we ran Botometer \cite{botOrNot} on the users that were removed from the mention graph, and found that their bot score was on average much lower than the remaining users (e.g. for the Princess dataset, the score was 0.97 vs 1.62, out of 5). 

    \item The best results, across all datasets, were achieved by $TWCRN$ (both for the follower graph and the mention graph) with an average F1 score of 0.86. This NN does not take into account the social links, at least not explicitly (perhaps the NN learns them implicitly). This network is also completely agnostic to epidemiological models. A clue to that $TWCRN$ learns the network structure in some implicit way lies in the fact that $TWCRN_{SHUF}$ fails miserably in all datasets.
    As we've mentioned in the introduction, many works have pointed out that information spreads  differently in social networks than epidemics spread in the population. Thus, it may be that  $TWCRN$ is able to learn this more complicated model of infection, which eludes simple human intuition.
    
    \item Having said that, we see that the simple $MLE$ algorithm performs surprisingly well on all datasets, achieving an average F1 score of 0.78 on the followers graph dataset. The $MLE$ follows a very simple intuition - it learns the user's ``trends", whether the user takes a break between tweets, or is a chain-twitter. Note that the $MLE$ algorithm completely ignores the social links.
    
    \item $TWMN$ is the only of our algorithms that takes into account the social links. Indeed it outperforms $MLE$, with an average F1 score of 0.81, but its performance is just slightly better than $TWMN_{all-1}$ (at an average F1 score of 0.8), where again the social links are not explicitly served. Hence we may conclude that the NN is able to learn by itself the important social links, and does not need it as an explicit input. 
    
    \item $TWPN$ performs better than the MLE, and the same as $TWMN$, at an average F1 score of 0.81. The intuition behind $TWPN$ is similar to  $MLE$, with the difference that $TWPN$ is basically learning a logistic regression per user rather than estimating the probabilities by averaging (although the tanh function is used and not the logit).
    
    \item The two baseline models $ALO$ and $LT$ perform poorly, slightly improving over the random guess. There are several ways to explain these poor results. This may be attributed to the way we have constructed the classification task, which may not be suitable for such models. 
    Another possibility is that the influence weights $\alpha_{u,v}$ that are used in Eq.~\eqref{eq:p_ALO} and \eqref{eq:p_LT}, which we derived from the weights of $TWPN$, were not a good choice. 
\end{itemize}


\section{Discussion}
In this work we checked two premises, the first is whether one can develop a prediction model which is completely agnostic to the motivation of information cascades as viewed from an epidemiological standpoint. If yes, then how much does the social link structure is a vital input to that prediction algorithm, or, can it be learned implicitly. We answered both questions positively, showing that a high F1 score is obtainable with a neural network that does not receive the social link structure explicitly, but does make use of it (as the random shuffle test shows). We also show that a slightly lower F1 score can be obtained by two simple algorithms, $MLE$ and $TWPN$, that completely ignore the social links. The role of the social links in predicting user activity should be further studied in future work, in other OSNs, and in other settings.

If we take out the Beirut dataset, which was our smallest dataset, then we discover that the ranking of F1 scores is inverse to the ranking of broadcasticity score, Princess is leading with highest F1 score and lowest broadcasticity, and Kobe vice-a-versa. 

We found that the results are significantly better in followers graph dataset, obtained by removing users that interact only with users that are not in their followers or friends list. Our first guess that these users are bots turned out false, and we leave this point as well for future work.

Finally, let us discuss the limitations of our approach. Our prediction was done for datasets that were collected around a single topic. We have not checked how well this method performs for a database that is not topic oriented. Our prediction task is time-constrained, namely, the input to the prediction algorithm is the activity in the last 12h, and prediction is made for the following 12h. While we played with these time slices and note a minor effect on the results, we did not try other configurations such as training for a week and predicting for the next day. We did not evaluate our models on platforms other than Twitter. Nevertheless, we expect them to generalize well on any platform which provides the same mechanisms for communication and social interaction since we did not use any domain-specific features such as linguistic features or other meta-data features.

\printbibliography

@article{galuba2010outtweeting,
  title={Outtweeting the twitterers-predicting information cascades in microblogs.},
  author={Galuba, Wojciech and Aberer, Karl and Chakraborty, Dipanjan and Despotovic, Zoran and Kellerer, Wolfgang},
  journal={WOSN},
  volume={10},
  pages={3--11},
  year={2010}
}

@inproceedings{petrovic2011rt,
  title={Rt to win! predicting message propagation in twitter},
  author={Petrovic, Sasa and Osborne, Miles and Lavrenko, Victor},
  booktitle={Proceedings of the International AAAI Conference on Web and Social Media},
  volume={5},
  number={1},
  year={2011}
}

@book{fine2006feedforward,
  title={Feedforward neural network methodology},
  author={Fine, Terrence L},
  year={2006},
  publisher={Springer Science \& Business Media}
}

@inproceedings{he2016deep,
  title={Deep residual learning for image recognition},
  author={He, Kaiming and Zhang, Xiangyu and Ren, Shaoqing and Sun, Jian},
  booktitle={Proceedings of the IEEE conference on computer vision and pattern recognition},
  pages={770--778},
  year={2016}
}

@inproceedings{matsubara2012rise,
  title={Rise and fall patterns of information diffusion: model and implications},
  author={Matsubara, Yasuko and Sakurai, Yasushi and Prakash, B Aditya and Li, Lei and Faloutsos, Christos},
  booktitle={Proceedings of the 18th ACM SIGKDD international conference on Knowledge discovery and data mining},
  pages={6--14},
  year={2012}
}

@inproceedings{tang2009relational,
  title={Relational learning via latent social dimensions},
  author={Tang, Lei and Liu, Huan},
  booktitle={Proceedings of the 15th ACM SIGKDD international conference on Knowledge discovery and data mining},
  pages={817--826},
  year={2009}
}

@inproceedings{kempe2003maximizing,
  title={Maximizing the spread of influence through a social network},
  author={Kempe, David and Kleinberg, Jon and Tardos, {\'E}va},
  booktitle={Proceedings of the ninth ACM SIGKDD international conference on Knowledge discovery and data mining},
  pages={137--146},
  year={2003}
}

@inproceedings{chen2010scalable,
  title={Scalable influence maximization for prevalent viral marketing in large-scale social networks},
  author={Chen, Wei and Wang, Chi and Wang, Yajun},
  booktitle={Proceedings of the 16th ACM SIGKDD international conference on Knowledge discovery and data mining},
  pages={1029--1038},
  year={2010}
}

@inproceedings{feng2014influence,
  title={Influence maximization with novelty decay in social networks},
  author={Feng, Shanshan and Chen, Xuefeng and Cong, Gao and Zeng, Yifeng and Chee, Yeow Meng and Xiang, Yanping},
  booktitle={Proceedings of the AAAI Conference on Artificial Intelligence},
  volume={28},
  number={1},
  year={2014}
}

@article{albert2002statistical,
  title={Statistical mechanics of complex networks},
  author={Albert, R{\'e}ka and Barab{\'a}si, Albert-L{\'a}szl{\'o}},
  journal={Reviews of modern physics},
  volume={74},
  number={1},
  pages={47},
  year={2002},
  publisher={APS}
}

@inproceedings{qiu2018deepinf,
  title={Deepinf: Social influence prediction with deep learning},
  author={Qiu, Jiezhong and Tang, Jian and Ma, Hao and Dong, Yuxiao and Wang, Kuansan and Tang, Jie},
  booktitle={Proceedings of the 24th ACM SIGKDD International Conference on Knowledge Discovery \& Data Mining},
  pages={2110--2119},
  year={2018}
}

@inproceedings{zhao2015seismic,
  title={Seismic: A self-exciting point process model for predicting tweet popularity},
  author={Zhao, Qingyuan and Erdogdu, Murat A and He, Hera Y and Rajaraman, Anand and Leskovec, Jure},
  booktitle={Proceedings of the 21th ACM SIGKDD international conference on knowledge discovery and data mining},
  pages={1513--1522},
  year={2015}
}

@inproceedings{mishra2016feature,
  title={Feature driven and point process approaches for popularity prediction},
  author={Mishra, Swapnil and Rizoiu, Marian-Andrei and Xie, Lexing},
  booktitle={Proceedings of the 25th ACM international on conference on information and knowledge management},
  pages={1069--1078},
  year={2016}
}

@article{moreno2002epidemic,
  title={Epidemic outbreaks in complex heterogeneous networks},
  author={Moreno, Yamir and Pastor-Satorras, Romualdo and Vespignani, Alessandro},
  journal={The European Physical Journal B-Condensed Matter and Complex Systems},
  volume={26},
  number={4},
  pages={521--529},
  year={2002},
  publisher={Springer}
}

@inproceedings{cha2009measurement,
  title={A measurement-driven analysis of information propagation in the flickr social network},
  author={Cha, Meeyoung and Mislove, Alan and Gummadi, Krishna P},
  booktitle={Proceedings of the 18th international conference on World wide web},
  pages={721--730},
  year={2009}
}

@article{delre2010will,
  title={Will it spread or not? The effects of social influences and network topology on innovation diffusion},
  author={Delre, Sebastiano A and Jager, Wander and Bijmolt, Tammo HA and Janssen, Marco A},
  journal={Journal of Product Innovation Management},
  volume={27},
  number={2},
  pages={267--282},
  year={2010},
  publisher={Wiley Online Library}
}

@inproceedings{lerman2010information,
  title={Information contagion: An empirical study of the spread of news on digg and twitter social networks},
  author={Lerman, Kristina and Ghosh, Rumi},
  booktitle={Proceedings of the International AAAI Conference on Web and Social Media},
  volume={4},
  number={1},
  year={2010}
}

@inproceedings{ver2011stops,
  title={What stops social epidemics?},
  author={Ver Steeg, Greg and Ghosh, Rumi and Lerman, Kristina},
  booktitle={Proceedings of the International AAAI Conference on Web and Social Media},
  volume={5},
  number={1},
  year={2011}
}

@article{goel2016structural,
  title={The structural virality of online diffusion},
  author={Goel, Sharad and Anderson, Ashton and Hofman, Jake and Watts, Duncan J},
  journal={Management Science},
  volume={62},
  number={1},
  pages={180--196},
  year={2016},
  publisher={INFORMS}
}

@inproceedings{botOrNot,
author = {Davis, Clayton Allen and Varol, Onur and Ferrara, Emilio and Flammini, Alessandro and Menczer, Filippo},
title = {BotOrNot: A System to Evaluate Social Bots},
year = {2016},
isbn = {9781450341448},
publisher = {International World Wide Web Conferences Steering Committee},
address = {Republic and Canton of Geneva, CHE},
url = {https://doi.org/10.1145/2872518.2889302},
doi = {10.1145/2872518.2889302},
booktitle = {Proceedings of the 25th International Conference Companion on World Wide Web},
pages = {273–274},
numpages = {2},
location = {Montr\'{e}al, Qu\'{e}bec, Canada},
series = {WWW '16 Companion}
}

@inproceedings{li2017deepcas,
  title={Deepcas: An end-to-end predictor of information cascades},
  author={Li, Cheng and Ma, Jiaqi and Guo, Xiaoxiao and Mei, Qiaozhu},
  booktitle={Proceedings of the 26th international conference on World Wide Web},
  pages={577--586},
  year={2017}
}

@article{kingma2014adam,
  title={Adam: A method for stochastic optimization},
  author={Kingma, Diederik P and Ba, Jimmy},
  journal={arXiv preprint arXiv:1412.6980},
  year={2014}
}

@article{cheng14,
author = {Cheng, Justin and Adamic, Lada and Dow, Alex and Kleinberg, Jon and Leskovec, Jure},
year = {2014},
month = {03},
pages = {},
title = {Can Cascades be Predicted?},
journal = {WWW 2014 - Proceedings of the 23rd International Conference on World Wide Web},
doi = {10.1145/2566486.2567997}
}

@inbook{difOfInov,
author = {García-Avilés, Jose},
year = {2020},
month = {09},
pages = {1-8},
title = {Diffusion of Innovation},
doi = {10.1002/9781119011071.iemp0137}
}

@inproceedings{ChengKLLSSA18,
  author    = {Justin Cheng and
               Jon M. Kleinberg and
               Jure Leskovec and
               David Liben{-}Nowell and
               Bogdan State and
               Karthik Subbian and
               Lada A. Adamic},
  title     = {Do Diffusion Protocols Govern Cascade Growth?},
  booktitle = {Proceedings of the Twelfth International Conference on Web and Social
               Media, {ICWSM} 2018, Stanford, California, USA, June 25-28, 2018},
  pages     = {32--41},
  publisher = {{AAAI} Press},
  year      = {2018},
  url       = {https://aaai.org/ocs/index.php/ICWSM/ICWSM18/paper/view/17876},
  bibsource = {dblp computer science bibliography, https://dblp.org}
  }

@article{WU2004327,
title = "Information flow in social groups",
journal = "Physica A: Statistical Mechanics and its Applications",
volume = "337",
number = "1",
pages = "327 - 335",
year = "2004",
issn = "0378-4371",
doi = "https://doi.org/10.1016/j.physa.2004.01.030",
url = "http://www.sciencedirect.com/science/article/pii/S0378437104000548",
author = "Fang Wu and Bernardo A. Huberman and Lada A. Adamic and Joshua R. Tyler",
}

@article{DynamicsOfViral,
author = {Leskovec, Jure and Adamic, Lada and Huberman, Bernardo},
year = {2005},
month = {10},
pages = {},
title = {The Dynamics of Viral Marketing},
volume = {1},
journal = {ACM Transactions on the Web},
doi = {10.1145/1134707.1134732}
}

@inproceedings{goel12,
author = {Goel, Sharad and Watts, Duncan and Goldstein, Daniel},
year = {2012},
month = {06},
pages = {},
title = {The structure of online diffusion networks},
journal = {Proceedings of the ACM Conference on Electronic Commerce},
doi = {10.1145/2229012.2229058}
}

@inproceedings{bakshy11,
author = {Bakshy, Eytan and Hofman, Jake and Mason, Winter and Watts, Duncan},
year = {2011},
month = {01},
pages = {65-74},
title = {Everyone's an Influencer: Quantifying Influence on Twitter},
journal = {Proceedings of the 4th ACM International Conference on Web Search and Data Mining, WSDM 2011},
doi = {10.1145/1935826.1935845}
}

@inproceedings{bergstra2011algorithms,
  title={Algorithms for hyper-parameter optimization},
  author={Bergstra, James and Bardenet, R{\'e}mi and Bengio, Yoshua and K{\'e}gl, Bal{\'a}zs},
  booktitle={25th annual conference on neural information processing systems (NIPS 2011)},
  volume={24},
  year={2011},
  organization={Neural Information Processing Systems Foundation}
}

@article{PA,
author = {Crammer, Koby and Dekel, Ofer and Keshet, Joseph and Shalev-Shwartz, Shai and Singer, Yoram},
year = {2006},
month = {03},
pages = {551-585},
title = {Online Passive-Aggressive Algorithms},
volume = {7},
journal = {Journal of Machine Learning Research}
}

@inproceedings{CascadesLSTM,
author = {Horawalavithana, Sameera and Skvoretz, John and Iamnitchi, Adriana},
year = {2020},
month = {04},
pages = {},
title = {Cascade-LSTM: Predicting Information Cascades using Deep Neural Networks}
}

@inproceedings{Lerman11digg,
author = {Ghosh, Rumi and Lerman, Kristina},
title = {A Framework for Quantitative Analysis of Cascades on Networks},
year = {2011},
isbn = {9781450304931},
publisher = {Association for Computing Machinery},
address = {New York, NY, USA},
booktitle = {Proceedings of the Fourth ACM International Conference on Web Search and Data Mining},
pages = {665–674},
numpages = {10},
}

@article{bass1969new,
  title={A new product growth for model consumer durables},
  author={Bass, Frank M},
  journal={Management science},
  volume={15},
  number={5},
  pages={215--227},
  year={1969},
  publisher={INFORMS}
}

@article{kermack1927contribution,
  title={A contribution to the mathematical theory of epidemics},
  author={Kermack, William Ogilvy and McKendrick, Anderson G},
  journal={Proceedings of the royal society of london. Series A, Containing papers of a mathematical and physical character},
  volume={115},
  number={772},
  pages={700--721},
  year={1927},
  publisher={The Royal Society London}
}

@article{hinton2006reducing,
  title={Reducing the dimensionality of data with neural networks},
  author={Hinton, Geoffrey E and Salakhutdinov, Ruslan R},
  journal={science},
  volume={313},
  number={5786},
  pages={504--507},
  year={2006},
  publisher={American Association for the Advancement of Science}
}


\appendix

\end{document}